# Can Alice and Bob be random: a study on human playing zero knowledge protocols.

KAMIL KULESZA

Department of Applied Mathematics and Theoretical Physics, University of Cambridge, Cambridge, UK[1],
Institute of Fundamental Technological Research, Polish Academy of Sciences, Warsaw, Poland,
e-mails: *K.Kulesza@damtp.cam.ac.uk, Kamil.Kulesza@ippt.gov.pl*

*Extended abstract.*

The research described in this abstract was initiated by discussions between the author and Giovanni Di Crescenzo in Barcelona in early 2004. It was during Advanced Course on Contemporary Cryptology that Di Crescenzo gave a course on zero knowledge protocols (ZKP), see [1]. After that course we started to play with unorthodox ideas for breaking ZKP, especially one based on graph 3-coloring. It was chosen for investigation because it is being considered as a "benchmark" ZKP, see [2], [3]. At this point we briefly recall such a protocol's description.

### *Graph 3-coloring (3COL) based ZKP*

*Input:* graph $G$

Alice (Prover) wants to prove to Bob (Verifier) she knows assignment of colors from the set $\{1,2,3\}$, such that it is a proper vertex 3-coloring of $G$. In addition, let $m = |E|$.

*One round of the protocol:*
1. Alice finds a random permutation $\varphi$ of the set $\{1,2,3\}$, which she applies to the coloring of $G$. Alice commits separately to coloring of each vertex and sends commitments together with the mapping into $G$'s vertices to Bob.
2. Bob selects at random an edge from $G$ and sends to it to Alice.
3. Alice responds with opening commitments at the ends of the edge selected by Bob.
4. Bob checks whether the vertices at the ends of the edge have different colors from the set $\{1,2,3\}$. If result is positive Bob accepts the round of the protocol, otherwise whole protocol is terminated and its result is negative.

The protocol is executed for $m^2$ rounds.

*Discussion.*

If Alice knows 3-coloring of $G$, she can convince Bob in every round of the protocol. If she does not know 3-coloring, she convinces Bob with probability $1 - \frac{1}{m}$, since for every round that must be at least one edge that vertices have same color (or one of the colors is from outside the set $\{1,2,3\}$). After $m^2$ rounds probability of Alice convincing Bob is $\left(1 - \frac{1}{m}\right)^{m^2} \approx e^{-m}$. Graph 3-coloring ZKP is a computational zero-knowledge proof, see [4]. ∎

A ZKP is usually performed between the machines. One of interesting questions is what happens if a human is made part of such a protocol. Apart from a theoretical considerations there is a least one hint that this might become more and more popular situation. The theorem by Manuel Blum states that any statement with a finite proof formulated in the logical proof system (eg. Russell and Whitehead system) can be proven by ZKP. He even entitled his original paper as "How to prove a theorem so no one else can claim it"([5]). So, what happens if people are allowed to play as Alice and Bob ?

---

[1] Part of the work described in this paper was done when the author was visiting scholar at DAMTP.

Below we outline theoretical reasoning and briefly describe practical experiments that took place.

First let's consider Bob being a human. His job is simple: chose an edge at random and check coloring of vertices. Any chance of cheating him?

Well, if Alice (Prover) can spot a pattern in Bob's choices for edges in the protocol rounds, she can employ adaptive strategy that increases chances for successful cheating. It is interesting as a way to attack ZKP from the Alice side, yet it is only a part of a larger problem, which we describe next.

Let's consider Alice being a human. She has twofold assignment: to convince Bob and to protect a secret, in this case 3-coloring of the graph. The later can be in danger if she does not permute colors at random. While the attack is a little tricky and beyond the scope of this paper, it should be emphasized that such patterns can be observed and leak information in certain protocol rounds. Obtained information allows to derive coloring for some parts of the graph, which with sufficient number of rounds can allow to recover coloring for the whole structure.

So, it should be clear by now that question of humans taking part in ZKP reduces very much to the quality of randomness that they can generate. Or more precisely how random they are when repeatedly choosing elements from a set of some fixed cardinality/fixed size?

In summer 2005, in order to answer this question we ran few series of experiments involving over 20 selected volunteers. They were mainly graduate and PhD students majoring in mathematics, computer science or physics. We also had few people with PhD from one of listed above fields. Test were carried out in computer lab with a help from custom made software.

We simulated ZKP by asking them generate random permutations, very much in a way like Alice in the real protocol. The experiment consisted of 4 different tests run on different days. Apart from permutations on 3 elements that correspond to 3COL, each person was asked to perform permutations on 2 and 4 elements. They were not forbidden to use any personal helping device. Also in about 50% of cases (chosen at random), individual was allowed to see list of permutations that she entered previously.

*Tests 1* and *Test 2* were designed as an introductory phase, which allowed to:
  a. get people used to the experiment;
  b. collect individual benchmark for every test participant;
  c. check whether experiment was executed correctly.

*Test 3* was proper simulation of 3COL ZKP (few hundred rounds), with an award promised to the best (the most random) performer. After the *Test 3* results were discussed with participants and emerging patterns presented. Only then participants learned that they were playing 3COL ZKP – the original instruction asked them to generate random permutations for "cryptographic purposes". Next, we held a seminar devoted to 3COL ZKP. Apart from the theory people participating in the experiment learned how patterns in generated permutations can compromise the secret. With all these knowledge and motivation they were asked to do *Test 4*, which setup was the same as *Test 3*. Next, results were analyzed across all tests. The analysis was carried out for every individual concerned as well as on aggregated results.

The picture varied much from person to person, allowing to compute individual fingerprints. Few individual obtained results of quality superior to others. As far as we know they were not using anything more then their brains.

Aggregated results were biased to the extend which allows to design general attack on 3COL ZKP with human Prover. It came as a surprise that results are strongly correlated with number of elements in the permutation.

We also run limited number of tests with people having less hard science exposure to see whether the results would differ much. Obtained results were so different that we almost started to think about people with background in hard science as different species.

In the paper we will discuss all obtained results in detail and present general conclusions.

Their applicability extend far beyond zero knowledge proof, with serious consequences to human computer interaction, especially in the field of security protocols.

Still there is a lot questions to be investigated, a chance to discuss them at Security Protocols Workshop would be of great value.